\documentclass[12pt,a4paper]{article}

\parskip=\baselineskip

\usepackage{hyperref}
\usepackage{graphicx}
\usepackage{color}
\usepackage{amsmath}
\usepackage{amssymb}
\usepackage[utf8]{inputenc}

\def\be#1\ee{\begin{align}\begin{split}#1\end{split}\end{align}}
\def\beq#1\eeq{\begin{align}\begin{split}#1\end{split}\end{align}}

\newcommand{\mysection}[1]{\section{#1}}
\newcommand{\mysubsection}[1]{\subsection{#1}}
\newcommand{\conjecture}[1]{\medskip\noindent{\bf #1}}

\newcommand{\mS}{m}

\begin{document}

\def\abstracttext{A low-energy effective theory is said to be in the swampland if it does not have any consistent UV completion inside a theory of quantum gravity. The natural question is if the standard model of particle physics, possibly with some minimal extensions, are in the swampland or not. We discuss this question in view of the recent swampland conjectures. We prove a no-go theorem concerning the modification of the Higgs sector. Moreover, we find that QCD axion is incompatible with the recent swampland conjectures, unless some sophisticated possibilities are considered. We discuss the implications of this result for spontaneous breaking of CP symmetry. We comment on dynamical supersymmetry breaking as well as the issue of multi-valuedness of the potential.}

\pagenumbering{Alph} 
\begin{titlepage}
\begin{flushright}
{\tt IPMU-18-0143}
\end{flushright}
\vskip 1in
\begin{center}
{\bf\Large{Do We Live in the Swampland?}}
\vskip 0.5cm 
{Hitoshi Murayama$^{1,2,3}$, Masahito Yamazaki$^1$ and Tsutomu T.~Yanagida$^1$} \vskip 0.05in 
\vskip 0.5cm 
{\small{ 
\textit{$^1$Kavli Institute for the Physics and Mathematics of the Universe (WPI), }\vskip -.4cm
\textit{University of Tokyo, Kashiwa, Chiba 277-8583, Japan}
\vskip 0 cm 
\textit{$^2$Berkeley Center for Theoretical Physics and Department of Physics,}\vskip -.4cm
\textit{University of California, Berkeley, CA 94720, USA}
\vskip 0 cm 
\textit{$^3$Physics Division, Lawrence Berkeley National Laboratory, }\vskip -.4cm
\textit{Berkeley,  CA 94720, USA }
}}

\end{center}

\vskip 0.5in
\baselineskip 16pt
\begin{abstract}
\abstracttext
\end{abstract}
\date{September, 2018}
\end{titlepage}
\pagenumbering{arabic} 

\mysection{Introduction}

String theory has been one of the most promising candidates for the theory of quantum gravity.
While string theory has been very successful in a number of different directions,
a fundamental question is if it has any direct experimental consequences in particle phenomenology.
The effective field theory dogma suggests that quantum gravity at the Planck scale
is irrelevant for a particle physicist, who often studies energy scales much lower than the Planck scale.

However, there is growing evidence that a vast class of effective field theories,
which are totally consistent as low-energy effective theories,  do not have consistent
UV completions with gravity included. In such cases, in the terminology of \cite{Vafa:2005ui,Ooguri:2006in}, 
the low-energy effective theories are in the {\it swampland} as opposed to the {\it landscape}.
Indeed, there are indications that a significant portion of the low-energy effective field theories fall into the swampland.
If this is indeed the case, it could  be misleading to be confined to the effective field theory framework:
the constraints for the existence of suitable UV completion with gravity imposes
important constraints for particle physics, which are hard to see otherwise.

It is then natural to ask the following question: is the standard model of particle physics, possibly with some extensions, 
be in the swampland or not? In this paper we discuss this question.
Our main focus will be the QCD axion,
which has long considered to be one of the most promising solutions 
to the strong CP problem \cite{Peccei:1977hh,Peccei:1977ur,Weinberg:1977ma,Wilczek:1977pj}.

To set the stage, we begin by summarizing the recent swampland conjecture of \cite{Obied:2018sgi}.
We then recall the quintessence explanation for the 
present-day vacuum energy \cite{Obied:2018sgi,Agrawal:2018own} (section \ref{section.Q}), and for the Higgs potential \cite{Denef:2018etk} (section \ref{section.Higgs}). We then come to our main ingredient, the QCD axion.  After pointing out the problem with QCD axion (section \ref{section.axion}), 
we discuss some possible loop holes (section \ref{section.loophole}). Our conclusion is that 
the QCD axion is excluded by a set of swampland conjectures \cite{Obied:2018sgi,Ooguri:2006in,ArkaniHamed:2006dz}, 
unless exotic scenarios are considered.
We discuss the implications of this result for the strong CP problem (section \ref{section.spontaneous}).
We comment on dynamical supersymmetry breaking and multi-valuedness of the potential (section \ref{section.remarks}).
We finally comment on the modification of the swampland conjecture (section \ref{sec.conjecture})
The Appendix (Appendix \ref{section.VH}) contains some no-go result for the modification of the Higgs potential.
The mathematical result there could be useful for an analysis of the conjecture in \cite{Obied:2018sgi} in other contexts.

\mysection{Swampland Conjectures}\label{sec.swampland}

Suppose that we have an effective field theory coupled with Einstein gravity, containing a finite number of scalar fields $\{\phi^i \}$.
We then have the Lagrangian
\be
\mathcal{L}=
\sqrt{-g} \left[ R+\sum_i D^{\mu} \phi^i D_{\mu} \phi^i + V_{\rm total}(\{ \phi_i \})+ \dots \right] \;.
\label{L}
\ee
Here $V_{\rm total}(\{ \phi_i \})$ is the potential for the scalar fields, and we added an index `total' to emphasize that this is the 
full potential for all the scalar fields in the theory. Note that for the scalar fields we have chosen a canonical kinetic term in the Einstein frame.
If this is not the case then a suitable re-parametrization of the fields is needed to bring the Lagrangian into the form of \eqref{L}.

The question is when this theory has a well-defined UV completion inside a suitable theory of quantum gravity, such as string theory.
In other words, is the theory in the swampland, or in the landscape?

In the literature several necessary conditions, for the effective theory to be in the landscape, have been proposed. We call these the swampland conjectures.
Over the years several such conjectures have been proposed \cite{Vafa:2005ui,Ooguri:2006in,ArkaniHamed:2006dz,Ooguri:2016pdq,Obied:2018sgi}, see \cite{Brennan:2017rbf} for recent summary.

One of the most recent of such swampland conjectures is the following remarkable conjecture due to 
Obied et.\ al.\ \cite{Obied:2018sgi}:

\conjecture{Conjecture 1} (de Sitter derivative conjecture): The potential $V_{\rm total}$ satisfies the following inequality:\footnote{Recall that the size of the gradient is given by 
\be
||\nabla V_{\rm total}||=\sqrt{\sum_{i,j} g_{\rm conf}^{ij} (\partial_{\phi_i} V_{\rm total} )(\partial_{\phi_j} V_{\rm total} )} \;,
\ee 
where $g_{\rm conf}^{ij}$ is the inverse metric on the configuration space for the scalar fields. In practice, in the following we always have a canonical diagonal metric $g_{\rm conf}^{ij}=\delta^{ij}$.}
\be
M_{\rm Pl}\, ||\nabla V_{\rm total} || > c_{\star}  V_{\rm total} \;,
\label{OV_conjecture}
\ee
where $c_{\star}$ is a $O(1)$ constant and $M_{\rm Pl}=2.4\times 10^{18} \textrm{ GeV}$ is the reduced Planck scale.

The precise optimal value of the $O(1)$ constant $c_{\star}$ depends on the setup. For example, four-dimensional compactifications of the eleven-dimensional supergravity suggests the value $c_{\star}=6/\sqrt{14}\sim 1.6$ \cite{Obied:2018sgi}. Our discussion below, however, does not depend on the precise values of $c_{\star}$, and could easily accommodate the value $c_{\star}\sim 10^{-3}$, for example.

We will discuss the phenomenological consequences of this and other swampland conjectures.
Note that even if the Conjecture 1 in itself does not hold in full generality, 
our conclusion still applies to some well-understood corners of string/M-theory vacua, 
as shown in the analysis of \cite{Hertzberg:2007wc,Wrase:2010ew,Obied:2018sgi}.
This is one of the reasons why the Conjecture 1 should be taken seriously.

\mysection{\bf Quintessence} \label{section.Q}

An immediate consequence of the conjecture \eqref{OV_conjecture} is that 
there is no de Sitter vacua, even metastable ones \cite{Danielsson:2018ztv,Obied:2018sgi} (cf.\ \cite{Maldacena:2000mw,McOrist:2012yc,Sethi:2017phn}):
\be
\quad V_{\rm total}>0 \;, \quad 
\nabla V_{\rm total}= 0   \quad \textrm{not allowed} \;.
\label{no_dS}
\ee
This in particular excludes the constant positive cosmological constant.
We can instead consider a dynamical vacuum energy as generated by a scalar field $Q$, the so-called quintessence field (\cite{Ratra:1987rm,Wetterich:1987fm,Zlatev:1998tr},
see \cite{Tsujikawa:2013fta} for review). 
This is an extremely light scalar field,
and we can for example choose the potential to be\footnote{Other potentials for quintessence has been proposed in the literature. Our conclusion does not 
depend much on the precise form of the quintessence potential, as long as the quintessence potential satisfies the constraint \eqref{OV_conjecture}.}
\be
V_Q(Q)=\Lambda_Q^4 \, e^{-c_Q \frac{Q}{M_{\rm Pl}}}  \;,
\label{V_Q}
\ee
where $c_Q$ is some $O(1)$ constant. It was shown that this kind of potential can indeed be incorporated in supergravity, namely the effective field theory of string theory \cite{Chiang:2018jdg}.

In this potential the shift of the origin of the quintessence field can be absorbed in the 
redefinition of the scale $\Lambda_Q^4$. We choose the present-day value of the 
quintessence field to be $Q/M_{\rm Pl}\sim 0$.
To explain the current value of the cosmological constant,
the energy scale $\Lambda_Q$ is chosen to be $\Lambda_Q \sim O(1) \textrm{ meV} $.\footnote{The possible connection that $\Lambda_Q \sim \Lambda_{\rm EW}^2/M_{\rm Pl}$ was explored in 
\cite{ArkaniHamed:2000tc}.}
The quintessence is the only scalar field at this energy scale,
and the condition \eqref{OV_conjecture} is satisfied easily if
$c_Q\ge c_{\star}$.

\mysection{Higgs}\label{section.Higgs}

In the minimal version of the standard model, the only fundamental scalar field is the Higgs particle.
At the electroweak (EW) scale $\Lambda_{\rm EW}\sim O(100) \textrm{ GeV}$, the only scalar fields in the theory
are the Higgs field $H$ and the quintessence field $Q$ introduced above.

The potential for the Higgs field is 
\be
V_H= \lambda (H^2-v^2)^2 \;,
\label{VH}
\ee
where $H$ is the absolute value of the complex Higgs field: in the following we always have the symmetry of rotating the phase of the complex Higgs field,
and the phase part of the Higgs field will not play any role. The total potential at the EW scale is then
\be
V_{\rm total}(Q,H)=V_Q(Q)+ V_H (H) \;.
\label{V_QH}
\ee
The Higgs potential \eqref{VH} has (a) a local minimum at $H^2=v^2$
and (b) a local maximum at $H= 0$. 

As already pointed out in \cite{Denef:2018etk}, the latter (namely the local maximum (b)) is contradictory with the swampland conjecture \eqref{OV_conjecture}.
In the neighborhood of the point (b) we obtain $\partial_H V_{\rm total}(Q,H)=\partial_H V_H(H)\sim 0$,
and hence $||\nabla V_{\rm total}(Q,H)||=\partial_Q V_Q(Q) \sim O(\Lambda_Q^4)$.
By contrast the value of the potential is given by  $V_{\rm total}(Q,H)\sim O(\Lambda_{\rm EW}^4)$
and is positive $V_{\rm total}(Q,H)>0$.
We therefore obtain
\be
0<M_{\rm Pl} \frac{||\nabla V_{\rm total}(Q,H)||}{V_{\rm total}(Q,H)} \sim O\left( \frac{\Lambda_Q^4}{\Lambda_{\rm EW}^4} \right)\sim O(10^{-56}) \;,
\ee
which is in sharp contradiction with the Conjecture 1 \eqref{OV_conjecture}.\footnote{This is the general structure when analyzing the Conjecture 1 in \eqref{OV_conjecture}---when the potential is a sum of contributions from several 
different energy scales with large hierarchies in between, then the existence of extremal values for the largest-energy-scale potential contradicts the conjecture \eqref{OV_conjecture}.}
Note that in this analysis we did not assume anything about the 
history of the Universe---the swampland conjecture applies to any possible field values which can be theoretically considered in the effective field theory.

One possibility to escape this contradiction is to modify the EW sector,
and consider a coupling of the Higgs field to some other field.
For example, we can introduce a real scalar field $S$ so that the 
potential is now given by
\be
V_{H,S}=\lambda (H^2-v^2)^2 + \kappa  (S-u) (H^2-w^2 ) + \frac{\mS }{2} S^2   +\Lambda_S^4 \;, 
\label{VHS}
\ee
where 
we have introduced new dimension-full parameters $u, w, \kappa, \mS , \Lambda_S$, which are assumed to be in the electroweak scale.
The potential \eqref{VHS} is the most general expression in $H^2$ and $S$ up to the quadratic order, up to a shift of the origin of $S$ and $H^2$. 
Since we have many free parameters, one might hope that one can adjust the free parameters such that 
there is no extremal values with positive potential values,
perhaps at the cost of fine-tuning the parameters.

It turns out, however, this modification does not work---one either finds a de Sitter extremum of the potential
and thus violating the conjecture \eqref{OV_conjecture}, or the EW vacua becomes unstable (the determinant of the Hessian about the EW vacua becomes negative).
The detailed analysis for this is provided in Appendix \ref{section.VH}. Indeed, we can show that this conclusion holds for a much broader class of models than
the particular model \eqref{VHS} (see a no-go theorem in Appendix \ref{section.VH}). 
While we did not completely exclude the possible EW modifications, we believe that this is a strong evidence that the EW modification of the Higgs sector requires more sophisticated scenarios, to say the least.

Instead of modifying the EW sector, we can take advantage of the quintessence field, as already pointed out in \cite{Denef:2018etk}.
Namely we can modify the Higgs potential to be 
\be
V' _H(H)=e^{-c_{H} \frac{Q}{M_{\rm Pl}}} \lambda' (H^2-v^2)^2 \;.
\label{HQ}
\ee
We can then easily verify that the combined potential $V_{\rm total}(H,Q)=V_Q(Q)+V' _H(H)$
does not have any extremal values.
We therefore no longer have any contradiction with the Conjecture 1 in \eqref{OV_conjecture}.

In conclusion, by applying the Conjecture 1 to the Higgs potential we obtained some supporting evidence for 
an existence of the quintessence field. This is independent from the argument from the previous section 
concerning the cosmological constant.

\mysection{QCD Axion}\label{section.axion}

In addition to the Higgs field, some extensions of the standard model
could contain other scalar fields, at energy scales lower than the EW scale.

A good example is the QCD axion,
which if present we will encounter
at the QCD scale $\Lambda_{\rm QCD}\sim O(100)\textrm{ MeV}$.
Since axions are abundant in string theory compactifications \cite{Witten:1984dg,Svrcek:2006yi}, it might be natural to imagine that 
we have the QCD axion in the string/M theory landscape.

The QCD axion, which we denote by $a$, couples to the QCD gauge field as a 
dynamical $\theta$-angle:
\be
\mathcal{L}_{\rm axion}=\frac{1}{32\pi^2} \frac{a}{f_a} e^{\mu\nu\rho \sigma} \textrm{Tr} F_{\mu\nu} F_{\rho \sigma}  ,
\ee
where $f_a$ is the axion decay constant.

Perturbatively we have a shift symmetry for the axion $a\to a+ (\textrm{const.})$, which is broken
only by the non-perturbative effects:
\be
V_{\rm axion}(a) =\Lambda_{\rm QCD}^4 \left[
 1-\cos \left(\frac{a}{f_a} \right) 
\right] \;.
\label{cosine}
\ee
This potential forces the axion to be at the origin,
and hence the QCD axion provides an elegant solution to the strong CP problem \cite{Peccei:1977hh,Peccei:1977ur,Weinberg:1977ma,Wilczek:1977pj}.

Let us consider the QCD scale where
only the quintessence and the axion are present, so that the total potential is given by
\be
V_{\rm total}(Q, a)=V_Q (Q) + V_{\rm axion}(a) \;.
\label{V_Q_phi}
\ee

There is a problem with the potential \eqref{V_Q_phi} at the field value $a=\pi f_a$,
which is a local maximum for the potential $V_{\rm axion}(a)$. We can now apply the similar logic as before:
while we have $||\nabla V_{\rm total}|| \sim |\partial_Q V_Q| \sim O(\Lambda_Q^4)$, the value of the potential is
given by $V_{\rm total}\sim V_{a} \sim O(\Lambda_{\rm QCD}^4)>0$, leading to the ratio
\be
0<M_{\rm Pl} \frac{||\nabla V_{\rm total}|| }{V_{\rm total}} \sim \frac{\Lambda_Q^4}{\Lambda_{\rm QCD}^4} \sim 10^{-44}  \;,
\ee
in contradiction with the Conjecture 1 in \eqref{OV_conjecture}.

As a cautionary remark, the cosine potential in \eqref{cosine} is obtained by the one-instanton approximation,
and does not quite match the actual form of the potential, as was suggested by the chiral Lagrangian analysis long ago \cite{DiVecchia:1980yfw,Witten:1980sp}
and studied more in detail by recent works (e.g.\ \cite{diCortona:2015ldu,DiVecchia:2017xpu}). However,
while the detailed form of the potential is different from \eqref{cosine},
the crucial fact that we have a local maximum at values $a\sim O(f_a)$ stays the same, and 
we still have a contradiction mentioned above.\footnote{See however the discussion in section \ref{subsection.branches}.}
While this is a rather simple argument, this potentially invalidates the QCD axion and hence 
one should try to find loopholes to the argument, as we will do below.

\mysection{In Search for Loopholes}\label{section.loophole}

One can think of several possible loopholes in the 
no-go discussion for the QCD axion in the previous section.
Let us discuss these in turn.

\mysubsection{Large Field Value}\label{subsection.large}

The first possible loophole is to make the value of the decay constant $f_a$ to be large,
so that we have
\be
f_a \gtrsim O(M_{\rm Pl}) \;.
\label{f_upper}
\ee

We can then appeal to the following swampland conjecture by Ooguri and Vafa \cite{Ooguri:2006in} (see also \cite{Grimm:2018ohb,Heidenreich:2018kpg,Blumenhagen:2018hsh,Lee:2018urn} for recent discussions):

\conjecture{Conjecture 2} (field range conjecture):
The range $\Delta \phi$ traversed by scalar fields in field space is bounded as $\Delta \phi \lesssim O(M_{\rm Pl})$;
at the field range $\Delta \phi\sim O(M_{\rm Pl})$ we inevitably encounter 
an infinite tower of nearly massless particles, thus invalidating the effective field theory.

When we assume the inequality \eqref{f_upper},
Conjecture 2 means that 
we have moved away the problematic value $a=\pi f_a$ beyond the regions of the 
validity of the effective field theory, thus removing the immediate contradiction with the Conjecture 1.

However, this is in sharp tension with yet another swampland conjecture, namely the 
weak gravity conjecture by Arkani-Hamed et.\ al.\ \cite{ArkaniHamed:2006dz} (see also \cite{Heidenreich:2015nta,Heidenreich:2016aqi,Harlow:2015lma,Cheung:2014vva,Montero:2016tif}). One particular consequence of the 
weak gravity conjecture is an existence of the upper bound on the decay constant \cite{ArkaniHamed:2006dz}: 

\conjecture{Conjecture 3} (weak gravity conjecture):
There is an upper limit on the decay constant
\be
f_a \lesssim \frac{M_{\rm Pl}}{S_{\rm inst}}\;, 
\ee
where $S_{\rm inst}$ is the value of the instanton action.

For QCD axions we have $S_{\rm inst}\sim 10^{2}$, to obtain
\be
f_a\lesssim 10^{-2} M_{\rm Pl} \;.
\label{f_lower}
\ee
The two results \eqref{f_upper} and \eqref{f_lower} are clearly contradictory.
The option of making the decay constant large is therefore eliminated.\footnote{While 
there are attempts to evade the weak gravity constraints by N-flation \cite{Liddle:1998jc,Copeland:1999cs,Dimopoulos:2005ac}
or alignment \cite{Kim:2004rp}, the $O(10^2)$ gap between the two constraints \eqref{f_upper} and \eqref{f_lower} makes is rather difficult to fill in the gap. See \cite{Rudelius:2015xta,Montero:2015ofa,Brown:2015iha,Heidenreich:2015wga} for related discussion.}

\mysubsection{Higher-Dimensional Operator}

The next is to appeal to higher-dimensional operators.
In the discussion of the potential \eqref{cosine} the shift symmetry of the axion
is only an approximately symmetry which is broken by non-perturbative effects.
The shift symmetry can instead by broken by quantum gravity effects represented as higher-dimensional operators in the Lagrangian.

For example, suppose that the $U(1)$ Peccei-Quinn symmetry 
is broken to a discrete $\mathbb{Z}_n$ subgroup,
by the effect of the higher-dimensional operator.
We then expect the following extra contribution to the axion potential
\be
\delta V_{\rm axion}(a)=\lambda \frac{f_a^{n}}{M_{\rm Pl}^{n-4}} \cos \left( \frac{n(a+a_0)}{f_a} \right) \;,
\label{higher-dim}
\ee
where $n$ is an integer (such that this term is a dimension $n$-operator),
and $\lambda$ and $a_0$ are the continuous parameters.

One should notice, however, that the potential as well as the original potential \eqref{cosine} is still periodic in $a$ with period $2\pi f_a$.
This immediately implies that there is still a maximum of the axion potential somewhere in the region $a\in [-\pi f_a, \pi f_a]$,
again in contradiction with the Conjecture 1.

In fact, the location of the maximum of  the potential stays close to the value $a\sim \pi f_a$.
To see this, note that the combined axion potential $V'_{\rm axion}(a)=V_{\rm axion} (a) + \delta V_{\rm axion}(a)$
near the origin is given by
\be
V'_{\rm axion}(a)=\lambda \frac{f_a^{n}}{M_{\rm Pl}^{n-4}} \sin a_0 \frac{na}{f_a} +\frac{\Lambda_{\rm QCD}^4}{2f_a^2}a ^2+O(a^3) +\ldots \;,
\ee
and hence the minimum of the axion potential $V'_{\rm axion}(a)$
is no longer at the origin $a=0$, and rather at a non-zero value $a=a_{\star}$, where $a_{\star}$ is given by
\be
a_{\star}\sim\frac{1}{f_a}\frac{\lambda \frac{f_a^{n}}{M_{\rm Pl}^{n-4}}}{ \frac{\Lambda_{\rm QCD}^4}{f_a^2}}=f_a  \frac{\lambda \frac{f_a^{n}}{M_{\rm Pl}^{n-4}}}{\Lambda_{\rm QCD}^4} \;.
\ee
For a solution of the CP problem (the small effective theta angle $|\bar{\theta}|=|\theta +\arg\det(m)|<10^{-9}$ for quark mass matrix $m$),
one needs
\be
\frac{a_{\star}}{f_a}\sim \frac{\lambda \frac{f_a^{n}}{M_{\rm Pl}^{n-4}}}{\Lambda_{\rm QCD}^4 }\sim \frac{\delta V_{\rm axion}(a)}{V_{\rm axion}(a)}<10^{-9} \;.
\ee
The contribution from the higher-dimensional operator \eqref{higher-dim} is therefore much smaller than the original potential \eqref{cosine} (cf.\ \cite{Kamionkowski:1992mf}).

\mysubsection{Coupling to Quintessence}\label{subsection.coupling_Q}

Another possible loophole is to consider the coupling to the quintessence. This might be
natural possibility to consider, since the similar solutions works for the Higgs potential, 
as explained before around \eqref{HQ}.

There is a big difference for the QCD axions, however.
The potential for the axion \eqref{cosine} is determined by the non-perturbative instanton effects,
and there is no option of modifying the potential \eqref{cosine}, say by coupling the axion directly to 
quintessence---one would then break the shift symmetry of the axion,
and hence the axion will no longer provide a solution to the strong CP problem.

One can still try to couple the quintessence field to the kinetic term of the axion.
This keeps the shift symmetry of the axion, and hence the potential \eqref{cosine}.
The total Lagrangian is now
\be
\mathcal{L}_{\rm total}=f\left(\frac{Q}{M_{\rm Pl}}\right) \, \partial_{\mu} a \partial^{\mu} a +V_{\rm axion} (a) + \partial_{\mu} Q \partial^{\mu} Q +V_Q (Q)\;,
\label{L_fQA}
\ee
for some function $f(Q/M_{\rm Pl})$. Since this is no longer has the canonical kinetic term,
one should do the field redefinition. 
We can choose the transformation
\be
a\to \int \! dQ \frac{1}{\sqrt{f\left(\frac{Q}{M_{\rm Pl}} \right)}}
:= g(Q) \;, \quad Q\to a \;,
\ee
so that the Lagrangian afterwards is
\be
\mathcal{L}_{\rm total}= \partial_{\mu} Q \partial^{\mu} Q +V_{\rm axion} (g(Q)) + \partial_{\mu} a \partial^{\mu} a +V_Q (a)\;.
\ee
We can exchange the label of $Q$ and $a$, to bring the Lagrangian into the more familiar form:
\be
\mathcal{L}_{\rm total}= \partial_{\mu} a \partial^{\mu} a +V_{\rm axion} (g(a)) + \partial_{\mu} Q \partial^{\mu} Q +V_Q (Q)\;.
\ee
This computation shows that for the analysis of the Conjecture 1, the 
only practical effect of the function $f(Q/M_{\rm Pl})$ is the replacement of the argument $a$ of $V_{\rm axion}$ by $g(a)$.
Despite this change, the potential $V_{\rm axion}(g(a))$ still has maximum at $a=a_{\rm max}$ with $g(a_{\rm max})=\pi f_a$,
and hence we run into the same contradiction with the Conjecture 1 as before. 

The only potential caveat for this is to appeal to the
loophole of section \ref{subsection.large}.
Suppose that the 
function $g(a)$ is chosen such that 
\be
a_{\rm max} \gtrsim O(10^{2}) \pi f_a\;,
\label{a_max}
\ee
such that the condition $a_{\rm max}\gtrsim M_{\rm Pl}$ 
can be imposed
without contradicting the constraints from the weak gravity conjecture \eqref{f_lower}:
\be
f_a\lesssim O(10^{-2}) M_{\rm Pl} \;.
\label{f_lower_2}
\ee
If $f_a$ saturates the bound \eqref{f_lower_2} (where the constraint should
be the least severe), we need
\be
g(M_{\rm Pl})\sim O(10^{-2})M_{\rm Pl}  \;.
\label{g_constraint}
\ee

This scenario is not impossible. For example, we can choose
\be
f\left(\frac{Q}{M_{\rm Pl}}\right)=e^{2  c_{QA} \frac{Q}{M_{\rm Pl}} }\;,
\label{f_QA}
\ee
so that we have
\be
\frac{g(Q)}{M_{\rm Pl}}=\frac{1}{c_{QA}}  \left( 1-e^{- c_{QA} \frac{Q}{M_{\rm Pl}} }  \right)\;,
\ee
Then \eqref{g_constraint} can be satisfied for $c_{QA}\sim 10^2$.

The interaction \eqref{L_fQA} for  the function \eqref{f_QA}  includes a linear coupling
\be
\mathcal{L}_{\rm total} \supset 2 c_{QA} \frac{Q}{M_{\rm Pl}} \, \partial_{\mu} a \partial^{\mu} a \;.
\ee
When we have a large coefficient $c_{QA}\sim 10^2$, this violates the Born unitarity of the 
$Q+a\to Q+a $ scattering amplitude before arriving at the Planck scale.

Other than coupling the quintessence to the kinetic term of the axion,
yet another possibility then is to keep the form of the potential \eqref{cosine},
and make the parameter $\Lambda_{\rm QCD}$ dependent on the quintessence:
\be
V_{\rm axion}(Q, a) =\Lambda_{\rm QCD}(Q)^4 \left[
 1-\cos \left(\frac{a}{f_a} \right) 
\right] \;.
\label{Q_cosine}
\ee
This can indeed be realized by 
coupling the quintessence $Q$ to the kinetic term for the gluons:
\be
\mathcal{L}_{\rm kin.}= \left(1+\lambda_{QFF} \frac{Q} {M_{\rm Pl}}  \right) \frac{1}{2g^2} \textrm{Tr} F_{\mu\nu} F^{\mu\nu} \;.
\label{QFF}
\ee
This is equivalent to making the gauge coupling constant $Q$-dependent:
\be
\frac{1}{g^2} \to \frac{1}{g(Q)^2}:= \left(1+\lambda_{QFF} \frac{Q} {M_{\rm Pl}}  \right) \frac{1}{g^2} \;,
\ee
which leads to the $Q$-dependence of the QCD scale $\Lambda_{\rm QCD}$ after transmutation:
\be
\Lambda_{\rm QCD}(Q)^4 &=\Lambda_{\rm cutoff}^4 \exp\left( -\frac{8\pi}{b_0}\left(\frac{1}{ g(Q)^2}-i \frac{\theta}{8\pi^2} \right)\right)  \\
& \to \Lambda_{\rm cutoff}^4 \exp\left( -\frac{8\pi}{b_0}\left( \left(1+\lambda_{QFF} \frac{Q} {M_{\rm Pl}}  \right) \frac{1}{g^2} -i \frac{\theta}{8\pi^2} \right)\right)  \sim \exp\left( -c'_{Q} \frac{Q} {M_{\rm Pl}} \right) \;,
\label{Lambda_Q}
\ee
where
\be
c'_{Q}=\frac{8 \pi \lambda_{QFF} }{b_0 g^2} \;.
\ee
and $b_0$ is the coefficient of the one-loop beta function.
The constraint from the Conjecture 1 in \eqref{OV_conjecture} is then satisfied by choosing $c'_Q\ge c_{\star}$.

The coupling \eqref{QFF} causes a serious problem, however. Once the quintessence couples to the gluons, then 
the quintessence couples to the nucleons through the gluon loops, 
so that we generate an effective interaction 
\be
\mathcal{L}_{QNN}\sim \lambda_{QNN} \frac{\Lambda_{\rm QCD}}{M_{\rm Pl}} Q NN  \;,
\ee
where $N$ here stands for nucleons.
Since we expect the coefficient $\lambda_{QNN}$ to be of $\sim O(1)$,
the coefficient is $\lambda_{QNN} \Lambda_{\rm QCD}/M_{\rm Pl}\sim O(10^{-19})$ and this is in tension with the equivalence-principle constraints on fifth-force between the nucleons: $(\textrm{Yukawa})<O(10^{-24})$ \cite{Wagner:2012ui}.
This is in contrast with the case of the Higgs particle, where
the similar coupling \eqref{HQ} between the Higgs and the quintessence is less constrained due to suppression of 
the loop diagrams by Yukawa couplings and electroweak couplings \cite{Denef:2018etk}.

While this eliminates the coupling \eqref{QFF} between the quintessence and the gluon,
one can try to save the loophole by coming up with a more complicated, if exotic, scenario.
One idea is to use the mirrored copy of the QCD \cite{Rubakov:1997vp,Berezhiani:2000gh,Hook:2014cda,Fukuda:2015ana}.
Here we have two copies of the QCD, our original QCD and its mirror image.
There is no direct coupling between the two copies of QCD.
We assume that the quintessence field couples only to the mirror QCD as in \eqref{QFF}, but not
to the original QCD. 
One then obtains the potential
\be
V_{\rm axion}(Q, a) =\left( \Lambda'_{\rm QCD}(Q)^4 +\Lambda_{\rm QCD}^4 \right) \left[
 1-\cos \left(\frac{a}{f_a} \right) 
\right] \;,
\label{double_cosine}
\ee
where the  mirror QCD scale $\Lambda'_{\rm QCD}(Q)^4$ comes from the mirror QCD (see \eqref{Lambda_Q})
\be
\Lambda'_{\rm QCD}(Q){}^4 &= \Lambda'_{\rm QCD}{}^4 \exp\left( -c'_{Q} \frac{Q} {M_{\rm Pl}} \right)  \;,
\ee
 and another scale $\Lambda_{\rm QCD}^4$
from the original QCD.\footnote{In the potential \eqref{double_cosine} we need to make sure that 
the phases of the two cosine functions from the two copies of QCD match. One expects that this is possible by imposing the 
mirror symmetry between the two copies of QCD at the value $Q=0$.}

The potential \eqref{double_cosine} satisfies the constraints from Conjecture 1 in \eqref{OV_conjecture}. 
Indeed, the total potential is now given by
\be
V_{\rm total}(Q, a) = V_Q(Q)+ V_{\rm axion}(Q, a)  \;,
\label{total_Qa}
\ee
where $V_Q$ is the quintessence potential \eqref{V_Q}.
The derivatives of the axion potential are computed to be
\be
M_{\rm Pl} \, \partial_a V_{\rm axion}(Q, a) &=\frac{M_{\rm Pl}}{f_a} \left( \Lambda'_{\rm QCD}(Q)^4 +\Lambda_{\rm QCD}^4 \right) \sin \left(\frac{a}{f_a} \right) \;, \\
M_{\rm Pl}\, \partial_Q V_{\rm axion}(Q, a) &= c'_{Q}{}^4 \Lambda'_{\rm QCD}(Q)^4  \left[
 1-\cos \left(\frac{a}{f_a} \right) \right] \;.
\ee
We find that the problematic point $a=\pi f_a$ no longer 
extremizes the potential. We still have $a=0$ as an extremal point, but this is of course
the minimum $V_{\rm axion}\sim 0$ of the axion potential and at this point the total potential
\eqref{total_Qa},
as well as the norm of the gradient of the potential,
 is dominated by quintessence contribution $V_Q$, which satisfies the Conjecture 1
as discussed in section \ref{section.Q}.

There is no constraint from the long-range force since the quintessence does not 
couple to the original copy of the QCD.

\mysubsection{Higgs Revisited}

Suppose that we have managed to evade the constraints on the QCD axion,
so that the Conjecture 1 in \eqref{OV_conjecture} is satisfied for the total potential $V_{\rm total}(Q,a)=V_{Q,a}(Q,a):=V_Q(Q)+V_{\rm axion}(Q,a)$ at the QCD scale. Namely, we have
\be
M_{\rm Pl}\, \sqrt{\left( \partial_Q V_{Q,a}(Q,a) \right)^2+\left(\partial_a V_{Q,a}(Q,a)\right)^2} \sim V_{Q,a}(Q,a) \lesssim O(\Lambda_{\rm QCD}{}^4) \;.
\label{der_Qa}
\ee
for all possible values of $Q$ and $a$.

Since we now have the QCD axion,
we should re-do the analysis of the Higgs potential in section \ref{section.Higgs}
at the EW scale. Let us start with the standard Higgs potential \eqref{VH} (with no coupling to the quintessence field),
so that the total potential at the EW scale \eqref{V_QH} now includes the axion:
\be
V_{\rm total}(Q,a, H)=V_Q(Q)+V_{\rm axion}(Q,a)+ V_H (H) \;,
\label{V_QaH}
\ee
where $V_H$ is the standard Higgs potential \eqref{VH}.

Let us study the neighborhood of the local maximum $H=0$ of the Higgs potential,
where $V_H(H)\sim 0$. We then have, using \eqref{der_Qa},
\be
M_{\rm Pl} ||\nabla V_{\rm total}(Q,a, H)|| =M_{\rm Pl}\, \sqrt{\left(\partial_Q V_{Q,a}(Q,a)\right) ^2+\left(\partial_a V_{Q,a}(Q,a)\right)^2}\lesssim O(\Lambda_{\rm QCD}{}^4) \;.
\ee
This implies
\be
0<M_{\rm Pl} \frac{||\nabla V_{\rm total}(Q,a,H)||}{V_{\rm total}(Q,a, H)} \lesssim O\left( \frac{\Lambda_{\rm QCD}^4}{\Lambda_{\rm EW}^4} \right)\sim O(10^{-12}) \;.
\label{ratio}
\ee
This is still in contradiction with Conjecture 1 in \eqref{OV_conjecture}. We can eliminate this problem by the 
coupling of the quintessence to the Higgs potential \eqref{HQ}, as in section \ref{section.Higgs}.

There seems to be  a possible loophole in this argument. In the discussion above (e.g.\ in \eqref{der_Qa})
we have implicitly assumed that the QCD scale $\Lambda_{\rm QCD}$ is the only scale relevant for the QCD axion.
This is not the case when we have mirror copies of QCD as in \eqref{double_cosine},
where we also have the mirror QCD scale $\Lambda'_{\rm QCD}$.
This scale can taken to be $\Lambda'_{\rm QCD} \sim \Lambda_{\rm EW}$ \cite{Fukuda:2015ana},
in which case other ratio in \eqref{ratio}
will be replaced by an $O(1)$ constant. Namely, for $H=0$ and $a\nsim 0$ one finds
\be
0<M_{\rm Pl} \frac{||\nabla V_{\rm total}(Q,a,H)||}{V_{\rm total}(Q,a, H)} \sim O\left( \frac{\Lambda'_{\rm QCD}{}^4}{\Lambda_{\rm EW}^4} \right)\sim O(1) \;.
\label{ratio_2}
\ee

There is another problem in the neighborhood of the special locus $H=a=0$, however. In this special case
both the axion potential $V_{\rm axion}$ and the Higgs potential $V_H$ are extremized,
and the norm of the gradient of the potential is given by the quintessence potential $V_Q$, so that
$||\nabla V_{\rm total}(Q,a,H)|| \sim O(\Lambda_Q^4)$. By contrast the value of the potential is dominated by the 
Higgs contribution, so that we have $0<V_{\rm total}(Q,a,H) \sim O(\Lambda_{\rm EW}^4)$.
We therefore find 
\be
0<M_{\rm Pl} \frac{||\nabla V_{\rm total}(Q,a,H)||}{V_{\rm total}(Q,a, H)} \sim O\left( \frac{\Lambda_{Q}{}^4}{\Lambda_{\rm EW}^4} \right)\sim O(10^{-56}) \;,
\label{ratio_3}
\ee
which is again in contradiction with the Conjecture 1.

\mysection{Spontaneous CP Breaking}\label{section.spontaneous}

Having discussed the possible loopholes in the previous section, we now arrived at one of our main conclusions.
Let us assume the recent swampland conjecture (Conjecture 1 in \eqref{OV_conjecture}) as well as the 
two more swampland conjectures (Conjecture 2 and Conjecture 3 in section \ref{subsection.large}),
and of course impose observational constraints. Then in effective field theories 
admitting a consistent UV completion inside theories of quantum gravity,
almost all of the existing scenarios for the QCD axion are ruled out.\footnote{Our conclusion applies only to the QCD axions, and does not 
necessarily exclude more general non-QCD axions.}

One should quickly add that there are still existing scenarios which evades these constraints,
such as the possibility discussed towards the end of section \ref{subsection.coupling_Q}. 
Regardless of this, it seems fair to say that swampland conjecture seems
to disfavor QCD axions.

How should we interpret our findings? 

One possibility is the one of the swampland conjectures,
say the Conjecture 1 given in \eqref{OV_conjecture}, does not hold (see section \ref{sec.conjecture} for a related conjecture).
Whether or not this is the case has been the matter of active discussion,\footnote{The literature is too large to be 
summarized here. See \cite{Andriot:2018wzk,Andriot:2018wzk,Andriot:2018wzk,Aalsma:2018pll,Garg:2018reu,Roupec:2018mbn,Andriot:2018ept,Heisenberg:2018yae,Conlon:2018eyr,Dasgupta:2018rtp,Kachru:2018aqn,Cicoli:2018kdo,Akrami:2018ylq} for a sample of recent references which 
discuss the construction of de Sitter vacua in string theory.}
and it would be desirable to come to a definite conclusion in the near future.
Regardless of the outcome, let us emphasize again that the Conjecture 1 is known to hold 
in some corners of string/M-theory vacua.

Let us for now suppose that the swampland conjectures are true.
Then we sill need to solve the strong CP problems. There are
several options.

\begin{itemize}

\item One still uses the QCD axion. As mentioned already this requires some sophisticated model building, 
such as the possibility discussed towards the end of section \ref{subsection.coupling_Q}.

\item There has been a proposed solution of the QCD by making the up quark (nearly) massless \cite{Banks:1994yg}. 
This option seems to be disfavored by lattice gauge theories \cite{Aoki:2013ldr}, which suggests non-zero up quark mass
with high statistical significance; see \cite{Dine:2014dga,Frison:2016rnq} for recent discussion.

\item Another possibility is that the CP symmetry is an exact symmetry of the Lagrangian (so that the bare value of the theta angle is 
$\theta=0$), and that the CP symmetry is spontaneously broken. 
Such a scenario was consider before, see e.g.\ \cite{Nelson:1983zb,Barr:1984qx,Masiero:1998yi,Evans:2011wj}.
In view of the results of this paper, it would be interesting to study if 
any of these models can be properly embedded into string theory.\footnote{Perturbative analysis of Calabi-Yau compactifications show that the CP is either unbroken, or broken by the vacuum expectation value of the CP-odd moduli \cite{Strominger:1985it,Green:1987mn}.
Even non-perturbatively it believed that CP is a gauge theory in string theory
and can be broken only spontaneously \cite{Dine:1992ya,Choi:1992xp}.}

\item Of course there could be other solutions of the strong CP problem, not traditionally discussed in the literature.
See the recent paper \cite{Cecotti:2018ufg} for such an attempt.

\end{itemize}

It is too early to tell which of these possibilities are realized in Nature. Regardless of the result, it is 
tantalizing that the insights from the  quantum gravity are now intimately tied with the
phenomenological search for the solutions of the strong CP problem.

\mysection{Beyond the Higgs}\label{section.remarks}

\mysubsection{Dynamical Supersymmetry Breaking}

In this paper we discussed the implications of the conjecture at the 
energy scales for the quintessence, QCD axion and the Higgs.
We can try to go further to higher energy scales.
While the analysis there depends on the details of the physics beyond the standard  model,
one ingredient one might wish to include is the supersymmetry, which we hope to be
broken dynamically \cite{Witten:1981nf}.

For some models of the dynamical supersymmetry breaking,
it is subtle to understand whether or not the Conjecture 1 excludes the model.
For example, in the Polonyi model, the conclusion depends crucially
on the behavior of the K{\" a}hler potential
when the Polonyi field takes an $O(1) M_{\rm Pl}$  value. This is also the case when 
the Polonyi model arises dynamically, as in the case of the IYIT model \cite{Izawa:1996pk,Intriligator:1996pu}, see \cite{Izawa:2008sa}.

The conclusion is much more clear-cut for other cases.
For example, in the models of metastable supersymmetry breaking
(such at the ISS model \cite{Intriligator:2006dd}), the supersymmetry is broken
at a  metastable de Sitter vacuum, where the field value is parametrically smaller than the Planck scale
and hence the physics is still calculable. The existence of such vacua immediately contradicts the Conjecture 1 
in \eqref{OV_conjecture}. This is an important consequence of the Conjecture 1---such metastable supersymmetry breaking is known to dramatically simplify the supersymmetric model building \cite{Murayama:2006yf,Murayama:2007fe}, but these are excluded by the 
swampland conjecture.

\mysubsection{Multi-Valuedness of the Potential and Inflation}\label{subsection.branches}

Let here us comment on one subtlety concerning the Conjecture 1.
In the formulation of the conjecture it is implicitly assumed that the potential is single-valued.
However, there are situations where the potential is multi-valued as a function of the field value,
say $\phi$.
In other words, we have several stable as well as metastable branches labeled by $1, 2, \dots$
with different potentials $V_1(\phi), V_2(\phi), \dots$, and we will have transitions between the branches.
In this case, the energy-minimizing potential, which corresponds to the stable branch,
is given by
\be
V_{\rm min}(\phi) = \min_{n} V_n(\phi) \;.
\label{V_min}
\ee

In this situation, we can consider two different possibilities in interpreting Conjecture 1:
\begin{enumerate}
\item We impose the constraint \eqref{OV_conjecture} for each branch, 
namely for each function $V_n(\phi)$.

\item We impose the constraint \eqref{OV_conjecture} for the minimum-energy potential
$V_{\rm min}(\phi)$.

\end{enumerate}

Our proposal is that we should choose the latter option. 

This has important consequences regarding the Conjecture 1. While one expects that the potential $V_n(\phi)$ to be a continuous function of the field $\phi$, the energy-minimizing potential $V_{\rm min}(\phi)$ is 
in general a {\it discontinuous} function of $\phi$, when the minimal branch changes from a branch $n$ to another branch $n'$.
This means the some mathematical results assuming continuity of the function,
such as the no-go theorem of Appendix \ref{app.theorem}, does not apply to the potential $V_{\rm min}(\phi)$.

Moreover, an inconsistency with the Conjecture 1 often happens when we have a local maximum of a smooth function,
which in our case is $V_n(\phi)$.  But when we have a local maximum in a branch $n$, then one might expect another branch 
where the value of the potential is smaller, so that the branch $n$ is not chosen for the energy-minimizing potential \eqref{V_min}.
One therefore expects that having the multi-branch structure will help to 
ameliorate the constraints from Conjecture 1.

An excellent example for such multi-branch structure is provided by an axion $a$ coupled with pure $SU(N)$ Yang-Mills
theory. In the large $N$ limit it was argued by Witten \cite{Witten:1980sp,Witten:1998uka} that we have an infinitely many branches 
labeled by an integer $n$, and the potential is given by
\be
V_n(a) =\frac{\Lambda_a^4}{2f_a^2} ( a-n\pi f_a)^2 \;, \quad
V_{\rm min}(\phi) = \min_{n\in \mathbb{Z}} \frac{\Lambda_a^4}{2f_a^2} ( a-n\pi f_a)^2 \;.
\label{largeN}
\ee
Note that the potential on each branch does not have the expected $2\pi f_a$ periodicity;
this periodicity is restored only after gathering together all the branches.
The potential \eqref{largeN} has discontinuities at $a=(2\mathbb{Z}+1) \pi f_a$,
where the sign of the derivative of the potential differs between the left and the right.
It is believed that the multi-branch structure is preserved even for a  finite value of $N$,
where we expect $O(N)$ branches of vacua (\cite{Witten:1980sp}, see \cite{Yamazaki:2017ulc,Aitken:2018mbb} for recent discussion). While the potential is 
quadratic near the origin, the potential eventually is bounded by the dynamical scale $O(1) \Lambda_a^4$,
and we expect a plateau near the values $a\sim N\pi f_a$.

The existence of such multi-branch structure was also observed in 
supersymmetric QCD \cite{Affleck:1983mk}. For the 
case of (non-supersymmetric) QCD, this was analyzed via the chiral Lagrangian in \cite{DiVecchia:1980yfw,Witten:1980sp} (see also \cite{Dashen:1970et}),
and we do have multiple branches for some quark masses.  We do not have such branch structures for realistic values of the quark masses, however. The discussion of section \ref{section.axion} is therefore not affected.

Let us finally comment on the relevance of this remark for inflation.\footnote{See e.g.\ \cite{Achucarro:2018vey,Kehagias:2018uem,Matsui:2018bsy,Garg:2018reu,Ben-Dayan:2018mhe,Kinney:2018nny} for recent discussion of the swampland conjectures in the context of inflation.} Instead of QCD axions we can choose the axion above to be the inflaton.
The multi-branch structure mentioned above gives a field-theory realization \cite{Kaloper:2011jz,Dubovsky:2011tu,Dine:2014hwa,Yonekura:2014oja,Kaloper:2016fbr} of the 
axion monodromy inflation, originally discussed in string theory \cite{Silverstein:2008sg,McAllister:2008hb}.

It has recently been pointed out that an inflation model based on this multi-branch structure \cite{Nomura:2017ehb} is in perfect agreement with the current observational constraints\footnote{The potential in \cite{Nomura:2017ehb} was inspired by the holographic computation of  \cite{Dubovsky:2011tu}, and is different from the cosine potential used for natural inflation \cite{Freese:1990rb}. Note that the deviation from the cosine potential for pure Yang-Mills is now firmly established by lattice gauge theory results \cite{Giusti:2007tu}, see also \cite{Nomura:2017zqj}.}. The inflaton rolls down the potential for a single branch ($V_n$ in the previous notation), since we can argue that 
the transition between different branches are irrelevant for the time scales of inflation \cite{Dubovsky:2011tu,Nomura:2017ehb}.
Since the model assumes the slow-roll condition, the current bounds for the scalar-to-tensor ratio
is in mild tension with the current Planck constraints ($c_{\star}\sim 0.1$ in \eqref{OV_conjecture}), as is the case in other slow-roll models \cite{Obied:2018sgi,Agrawal:2018own}.

It is worth pointing out that the setup of \cite{Nomura:2017ehb}, together with the proposed 
implementation of the Conjecture 1, eliminates the problem of the plateau of the inflaton potential.
In many of the inflationary models today the inflaton potential has a plateau region where the potential is nearly flat.
This is clearly a dangerous region for the Conjecture 1. Such a plateau, however, does not
appear in the energy-minimized potential $V_{\rm min}$ in \eqref{V_min}.
The multi-valued structure of the potential has traditionally been invoked for increasing the field range
traversed by the inflaton. What we are finding here is that it has a different virtue, namely the 
consistency with the swampland conjecture of \eqref{OV_conjecture}.  

\mysection{Modified Swampland Conjecture}\label{sec.conjecture}

In view of the phenomenological constraints discussed in this paper, 
one of the most natural possibilities is to weaken the swampland conjecture \eqref{OV_conjecture}.

One plausible possibility is to modify the conjecture \eqref{OV_conjecture} to be in the following form:
\footnote{See e.g.\ \cite{Andriot:2018wzk,Garg:2018reu} for other modification of the conjecture \eqref{OV_conjecture}.}
\be
M_{\rm Pl}\, ||\nabla V_{\rm total} || > c_{\star}  V_{\rm total}  \;, 
\quad \textrm{whenever}  \quad  \textrm{Hessian}( V_{\rm total}) > 0  \;.
\label{no_dS_2}
\ee
This should be compared with \eqref{no_dS}.
Namely, we allow for the point $V_{\rm total}>0, \nabla V_{\rm total}= 0$ as long as 
the Hessian has at least one non-positive eigenvalue. This restriction seems natural since
the point is unstable if the Hessian has a negative eigenvalue.
This proposal immediately removes the problem with the QCD axion and the Higgs
discussed in this paper.

\section*{Acknowledgements}

We would like to thank C.-I.~Chiang, H.~Fukuda, K.~Hamaguchi, A.~Hebecker, M.~Ibe, S.~Matsumoto, T.~Moroi, H.~Ooguri, S.~Sethi, S.~Shirai and C.~Vafa for discussions.
This research was supported in part by World Premier International Research Center Initiative, MEXT, Japan.
H.~M.\ was supported in part by U.S.\ DOE under Contract DE-AC02-05CH11231, NSF under grants PHY1316783 and PHY-1638509, JSPS Grant-in-Aid for Scientific Research (C) No.\ 26400241 and 17K05409, and MEXT Grant-in-Aid for Scientific Research on Innovative Areas No.\ 15H05887, 15K21733.
T.~T.~Y.\ was supported in part by the JSPS Grant-in-Aid for Scientific Research No.\ 26104001, No.\ 26104009, No.\ 16H02176, and No.\ 17H02878. T.~T.~Y.\ is a Hamamatsu Professor at Kavli IPMU.
M.~Y.\ was supported in part by the JSPS Grant-in-Aid for Scientific Research No.\ 17KK0087.

\appendix
\mysection{Analysis of the Higgs Potential}\label{section.VH}

As mentioned in the main text, one possible way to escape the constraint from the Conjecture 1 in \eqref{OV_conjecture}
is to extend the EW sector, so that we have a potential involving multiple fields.
In this Appendix we discuss some difficulties in this approach.

\mysubsection{The Potential of \eqref{VHS}}

Let us start with the potential of \eqref{VHS}, where we included a real field $S$ in addition to the Higgs field $H$.

By extremizing the potential ($\partial_H V_{H,S}=\partial_S V_{H,S}=0$), one finds two different solutions.
The first solution, which we call solution (a), is what should be the EW vacuum, corresponding to the value $H^2=v^2$ in the original Higgs potential \eqref{VH}:
\be
H_{\rm (a)}^2=\frac{\kappa ^2 w^2-\kappa  m^2 u-2 \lambda  m^2 v^2}{\kappa ^2-2 \lambda  m^2}\;, \quad
S_{\rm (a)}=\frac{\kappa ^2 u+2 \kappa  \lambda  v^2-2 \kappa  \lambda  w^2}{\kappa ^2-2 \lambda 
   m^2} \;. \\
\ee
Another solution, which we call solution (b), corresponds to the local maximum $H^2=0$ of the original Higgs potential \eqref{VH}:
\be  
H_{\rm (b)}^2=0 \;, \quad S_{\rm (b)}=\frac{w^2 \kappa}{\mS ^2} \;. 
\ee

There are several conditions to be imposed. 
First, since we wish to keep the EW vacuum (namely solution (a)), we need
\be
H_{(a)}^2=\frac{\kappa ^2 w^2-\kappa  m^2 u-2 \lambda  m^2 v^2}{\kappa ^2-2 \lambda  m^2} \ge 0 \;.
\label{condition1}
\ee

Second, we should have zero energy at the solution (a); if this is not the case we have non-zero constant cosmological constant
at lower energy scales, and we spoil the quintessence discussion in section \ref{section.Q}. This requires us to choose the constant $\Lambda_S^4$ to be
\be
\Lambda_S^4=-\frac{\kappa  \left(m^2 u \left(\kappa  u+4 \lambda  \left(v^2-w^2\right)\right)+2
   \kappa  \lambda  \left(v^2-w^2\right)^2\right)}{2 \left(\kappa ^2-2 \lambda 
   m^2\right)} \;.
\ee

Third, we impose the condition that the solution (a) is at least a local minimum.
This in particular implies that the determinant of the Hessian at (a) is positive, leading to the constraint
\be
m^2 \left(\kappa  u+2 \lambda  v^2\right)-\kappa ^2 w^2>0 \;.
\label{condition2}
\ee

Finally, for the consistency with the conjecture \eqref{OV_conjecture}
we require that the value of the potential is non-positive at the solution (b). This gives
\be
V_{(b)}=\frac{\left(\kappa ^2 w^2-m^2 \left(\kappa  u+2 \lambda  v^2\right)\right)^2}{4 \lambda 
   m^4-2 \kappa ^2 m^2} <0 \;,
\ee
namely
\be
\kappa ^2 w^2-m^2 \left(\kappa  u+2 \lambda  v^2\right)=0 \quad \textrm{or}\quad
\lambda<\frac{\kappa^2}{2 m^2} \;.
\label{condition3}
\ee

The three conditions \eqref{condition1}, \eqref{condition2}, \eqref{condition3} are mutually incompatible.
We therefore conclude that the potential \eqref{VHS} does not serve our purposes.

\mysubsection{General Possibilities: a No-Go Theorem}

While the discussion of the previous subsection was restricted to a particular potential \eqref{VHS},
the lesson is actually more general.

Suppose that 
we have a set of scalar fields $\vec{S}$ such that the total potential,
involving the Higgs field, is given by
\be
V_{H,S}(H, \vec{S}) =  V_H(H)+\dots,
\ee
where $\dots$ represents the terms involving the field $\vec{S}$.
We assume that $V_{H,S}$ is a continuous and differentiable function of the arguments $H$ and $\vec{S}$.

In general we find multiple solutions to the 
extremal condition: 
\be
\partial_H V_{H,S}=\partial_{\vec{S} }V_{H,S}=0 \;.
\label{extrem}
\ee

In general there are many other solutions to \eqref{extrem}, and 
Conjecture 1 in \eqref{OV_conjecture} could be violated at any of these points.
We therefor impose the condition
\begin{quote}
(A) The potential is non-positive at all the solutions of \eqref{extrem}.
\end{quote}

Moreover, we wish to have the EW vacuum ($H=v$ in the original Higgs potential \eqref{VH}).
This motivates us to impose 
\begin{quote}
(B) There exists a solution  of \eqref{extrem}, which is a local minimum for the potential $V_{H, S}$.
We moreover assume that there are no flat directions around the solution, and the value of the potential 
vanishes at the solution: $V_{H, S}=0$.
\end{quote}

Let us further assume that 
\begin{quote}
(C) There exists at least one solution to \eqref{extrem} other than the EW vacuum solution discussed in (A).
\end{quote}
Namely we exclude the possibility that (A) is the only extremal value of the potential in the configuration space.

It turns out that it is not possible to satisfy all the constraints (A), (B), (C).
This is our no-go theorem.

\mysubsection{Proof of the No-Go Theorem}\label{app.theorem}

Let us give a proof of this no-go theorem.\footnote{We thank Kyoji Saito for suggesting some refinement on this proof. The possible error in the following proof, however,
should be attributed solely to the authors.} 
Let us denote the EW vacuum of (B) as $P_B=(H_{\rm (B)}, S_{\rm (B)})$,
and another Anti-de Sitter vacuum of (C) as $P_C=(H_{\rm (C)}, S_{\rm (C)})$.
Let us assume we have a total of $D$ fields and the configuration space of $(H, \vec{S})$ is $D$-dimensional.

\begin{figure}[htbp]
\centering{\includegraphics[scale=0.3]{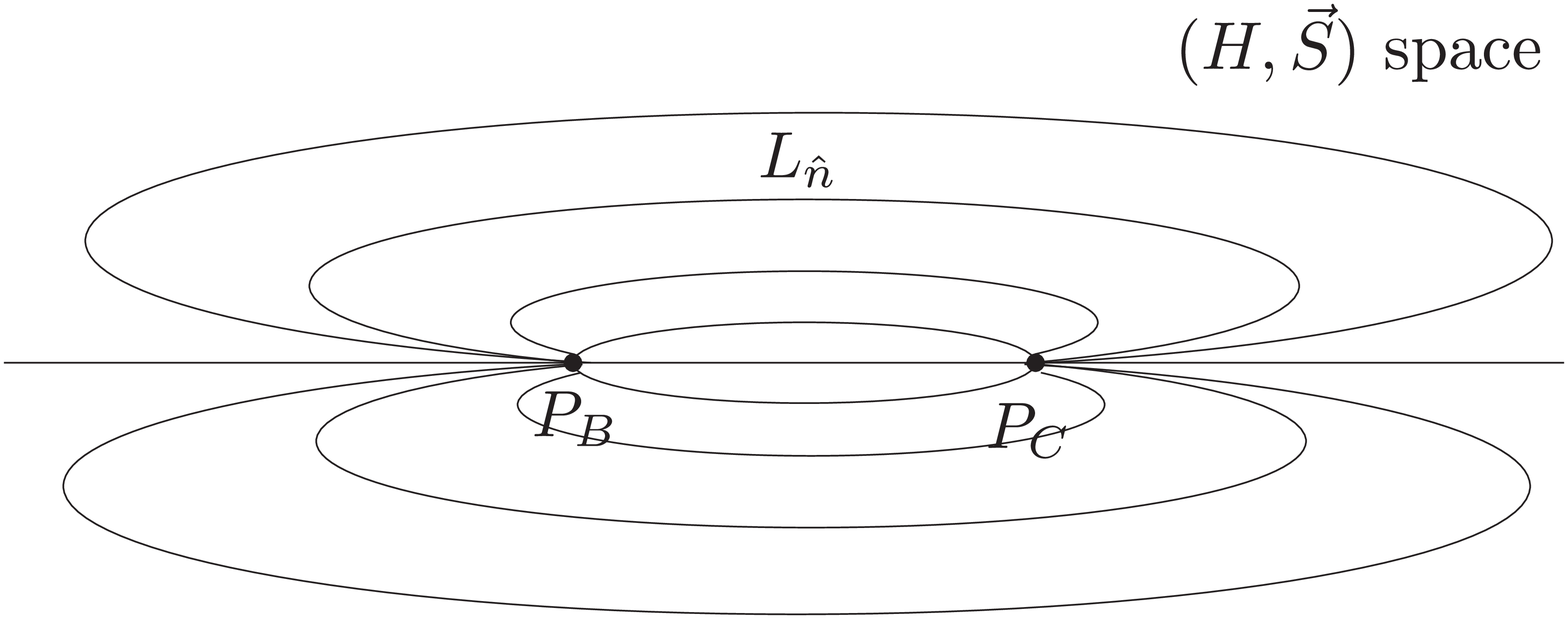}}
\caption{We can foliate the $D$-dimensional configuration space by a set of $L_{\hat{n}}$ with $\hat{n} \in S^{D-1}$
starting with $P_B$ and ending at $P_C$. Here we show the case of $D=2$, where $\hat{n}\in S^1$ is a point of the circle,
namely specifies the direction in the neighborhood of the point $P_B$.}
\label{fig.foliate}
\end{figure}

Let choose a set of path $L_{\hat{n}}$
starting from $P_B$ into $P_C$, so that (1) $L_{\hat{n}}$ points in the direction $\hat{n} \in S^{D-1}$ 
in the neighborhood of $P_B$ and then reaches $P_C$ and
(2) there are no mutual intersections of $L_{\hat{n}}$ with different $\hat{n} \in S^{D-1}$, so that
$L_{\hat{n}}$ with $\hat{n} \in S^{D-1}$  foliates the whole $(H, \vec{S})$-plane.
See Figure \ref{fig.foliate} for the case of $D=2$.
We can think of the combination $(P, \hat{n})$ with $P\in L_{\hat{n}}, \hat{n}\in S^{D-1}$
as providing a coordinate chart in the configuration space.

Let us fix $\hat{n} \in S^{D-1}$ and consider the function $V_{H,S}$ along the line
$L_{\hat{n}}$, starting with the point $P_B$. Since $P_B$ was the EW vacuum we start
with $V_{H,S}=0$, and by assumption (B) the potential grows into positive values
as we gradually move along $L_{\hat{n}}$. Since we know (by assumption (C) and (A)) that the potential should
reach negative values by the time we get to the point $P_C$,
and since the potential is the continuous function of the arguments,
we quickly conclude that there should be at least one local maximum 
along the path  $L_{\hat{n}}$. If there are multiple such local maximums, we take the 
one closest to the point $P_B$, and we call this point $P_{\hat{n}}$. Obviously we find 
$V(P_{\hat{n}})>0$.\footnote{Strictly speaking  $L_{\hat{n}}$ for some particular value  $\hat{n}_{\infty}$ of $\hat{n}$
runs off to infinity, one might worry that for this $\hat{n}=\hat{n}_{\infty}$ the path $L_{\hat{n}_{\infty}}$ has infinite length and the local maximum we mentioned here might be located at infinity. When this happens, we can replace the minimum in \eqref{def_star}
by a maximum to apply the same argument, to arrive at \eqref{VP_1} and \eqref{VP_2}.}

Let us now consider the values of the potential $V(P_{\hat{n}})$ as we change $\hat{n}\in S^{D-1}$.
Since $S^{D-1}$ is a compact space, there is necessarily a point $\hat{n}_*\in S^{D-1}$
which attains the minimum:
\be
V(P_{\hat{n}_*}) =\min_{\hat{n}\in S^{D-1}} V(P_{\hat{n}}) \;.
\label{def_star}
\ee

Since $V(P_{\hat{n}})>0$ for all $\hat{n}\in S^{D-1}$,
we in particular find that 
\be
V(P_{\hat{n}_*})>0 \;.
\label{VP_1}
\ee
Moreover, we find $P_{\hat{n}_*}$ is a extremal point of the potential:
\be
\nabla V(P_{\hat{n}_*})>0 \;.
\label{VP_2}
\ee
Indeed, the derivative of the potential vanishes
along the path $L_{\hat{n}_*}$ from the definition of $P_{\hat{n}_*}$,
and vanishes along the direction of the sphere $S^{D-1}$ thanks to the definition \eqref{def_star};
since the derivative of the potential vanishes in all the linearly-independent directions,
the derivative should vanish in all the directions.
The result \eqref{VP_1} and \eqref{VP_2} are in contradiction with our assumption.
This concludes our proof.

Our result excludes many of the possible EW modifications of the Higgs potential.
For example, if we have a polynomial potential $V_{H, \vec{S}}$ for 
complex scalars $\vec{S}$, then we generically expect many extremal points (thus satisfying (C)),
so that we can conclude without any explicit computations that the potential does not satisfy our criterion.
Note that the quintessence modification in \eqref{HQ} solves the problem by 
violating the condition (C). 

While we discussed this result in the context of the Higgs potential,
our mathematical no-go theorem can be used in other contexts, 
e.g.\ the discussion of the moduli stabilization in the the swampland conjecture (see \cite{Conlon:2018eyr} and version 2 of \cite{Denef:2018etk}
for one-parameter version of our discussion).

\bibliographystyle{nb}
\bibliography{swampland}
\end{document}